\documentstyle[12pt,a4fedele]{article}   

\newcommand{\be}{\begin{equation}}
\newcommand{\ee}{\end{equation}}
\newcommand{\bea}{\begin{eqnarray}}
\newcommand{\eea}{\end{eqnarray}}
\newcommand{\bean}{\begin{eqnarray*}}
\newcommand{\eean}{\end{eqnarray*}}

\def\thefiglist#1{\section*{Figure Captions\markboth
{FIGURE CAPTIONS} {FIGURE CAPTIONS}}\list
{Figure \arabic{enumi}.}
{\settowidth\labelwidth{Figure #1.}\leftmargin\labelwidth
\advance\leftmargin\labelsep
\usecounter{enumi}}
\def\newblock{\hskip .11em plus .33em minus -.07em}
\sloppy}

\begin{document}
\begin{flushright} RAL-92-026, CEBAF-TH-92-13, IU/NTC-92-16\\
\end{flushright}
\vskip 1.5cm
\centerline {\bf\large Scalar Mesons in $\phi$ Radiative Decay:}
\centerline {their implications for spectroscopy and for
studies of CP-violation at
$\phi$ factories}
\vskip 1.5cm
\centerline {\bf \large F. E. Close}
\centerline {Rutherford Appleton Laboratory}
\centerline {Chilton, Didcot, Oxon,
OX11 0QX, England.}
\vskip 0.5cm
\centerline  {\bf \large Nathan Isgur}
\centerline {CEBAF}
\centerline {12000 Jefferson Avenue,
Newport News, VA 23606, U.S.A.}
\vskip 0.5cm
\centerline { \bf \large S. Kumano}
\centerline {Nuclear Theory Center}
\centerline {Indiana University,\\
2401 Milo B. Sampson Lane, Bloomington, IN 47408, U.S.A.}
\bigskip
\bigskip
\bigskip
\centerline {\bf Abstract}

   Existing predictions for the branching ratio for
$\phi\rightarrow K\bar{K}\gamma$ {\it via} $\phi\rightarrow S\gamma$
(where $S$ denotes one of the scalar mesons $f_0$(975) and
 $a_0$(980)) vary by several orders of
magnitude. Given the importance of these processes for both
hadron spectroscopy and CP-violation studies
at $\phi$ factories (where
$\phi \rightarrow K^0 \bar K^0  \gamma$ poses a possible
background problem),
this state of affairs is very undesirable.
We show that the variety of predictions is due in part to errors
and in part to differences
in modelling. The latter variation leads us to argue that the
radiative decays
of these scalar states
are interesting in their own right
and may offer unique insights into the nature
of the scalar mesons.
As a byproduct we find that the branching ratio for
$\phi\rightarrow K^0\bar{K}^0\gamma$ is $\lapproxeq 0(10^{-7})$
and will pose
no significant background to proposed studies of CP-violation.

\vfill\eject

\section{Introduction}
{}~~

    There are predictions in
the existing literature for the branching ratio for
$\phi\rightarrow
K\bar{K}\gamma$ {\it via} $\phi\rightarrow S\gamma$ (where $S$ denotes
one of the scalar
mesons
 $S^*$ (now called $f_0(975)$) or $\delta$
(now called $a_0$(980)) that vary by several orders of
magnitude [1-5].  Clearly not all of these predictions
can be correct! Given the importance of these
processes for both hadron spectroscopy and CP-violation
studies, this state of affairs is clearly undesirable.
Moreover, in view of the impending $\phi$ factory, DA$\Phi$NE [6],
and other
developing programmes [7], there is an urgent need to clarify the
theoretical situation.

   The scalar mesons ({\it i.e.}, mesons with
$J^{PCn} =0^{++}$) have been a persistent problem
in hadron spectroscopy.\footnote{For an historical perspective see Ref.
[8]; for a more recent study see Ref. [9].}  We shall show in this paper
that the radiative
decays of the $\phi$ meson to these states can discriminate among various
models of their structure. In addition to the spectroscopic issues surrounding
the scalar mesons, there is a
significant concern that the decay
$\phi\rightarrow K^0\bar{K}^0\gamma$ poses a possible background
problem
to tests of CP-violation at future $\phi$ factories: the radiated photon
forces the
$K^0\bar{K}^0$ system to be in a C-even state, as opposed to the C-odd decay
$\phi\rightarrow K^0\bar{K}^0$. Looking for CP-violating
decays in $\phi\rightarrow
K^0\bar{K}^0$ has been proposed as a good way to
measure
$\varepsilon^\prime /\varepsilon$ [10],
but because this means looking for a small
effect, any
appreciable rate for $\phi\rightarrow K^0\bar{K}^0\gamma$
(namely, a branching ratio
$\phi\rightarrow
K^0\bar{K}^0\gamma\gapproxeq 10^{-6}$) will limit the precision of such an
experiment.
Estimates [4] of the {\it non-resonant} $\phi\rightarrow K^0\bar{K}^0\gamma$
rate give, in the absence of any resonant contribution, a
branching ratio of the order of  $10^{-9}$, far too
small to pose a problem. The uncertainty in the theoretical
estimates, and the potential experimental
ramifications, arise due to the presence
of the scalar mesons $f_0(975)$ and $a_0(980)$, which are strongly
coupled to the $K\bar{K}$ system.  Estimated rates for
the {\it resonant} decay chain
$\phi\rightarrow
S+\gamma$, followed by the decay $S\rightarrow
K^0\bar{K}^0$,
vary by three orders of magnitude, from
a branching ratio of the order of $10^{-6}$ down to $10^{-9}$. These variations
in fact reflect the uncertainties in the literature for
the expected branching ratio for $\phi\rightarrow S\gamma$ which vary from
$10^{-3}$ to $10^{-6}$ [11] .  Here
we concentrate on this resonant process.

  We shall show that the variability of the
predictions for $\phi \rightarrow S \gamma$ is due in part to errors
and in part to differences
in modelling. On the basis of this model
dependence, we argue that the study
of these scalar states in
 $\phi\rightarrow S\gamma$
may offer unique insights into the nature
of the scalar mesons. These insights should help
lead in the future to a better understanding of
not only quarkonium but also glueball
spectroscopy. As a byproduct we predict that the branching ratio for
 $\phi\rightarrow K^0\bar{K}^0\gamma$ is $\lapproxeq 0(10^{-7})$
({\it i.e.}, the branching ratio for $\phi\rightarrow
S\gamma$ is $ \lapproxeq 0(10^{-4}))$ and will pose
{\it no} significant background to studies of CP-violation at DA$\Phi$NE.

\section{Probing the nature of the scalar mesons below 1 GeV}
{}~~

The scalar mesons are spectroscopically
interesting for several reasons.  One is that,
while agreeing on little else, it is an essentially universal prediction of
theory (lattices, bags, flux tube models, QCD sum rules, \ldots ) that the
lowest-lying glueball has scalar quantum numbers and a mass in the 1.0
- 1.5 GeV mass range. Clarifying the presently confused nature of the known
$0^{++}$ mesons may be pivotal in the quest to identify this glueball.  Another
is
the possibility that the two best known [12] scalar mesons,
the $f_0$(975) and the $a_0$(980), are $qq\bar{q}\bar{q}$
states.  The original proposal [13] for this interpretation, based on the
bag model, also predicted many other states which have not been seen
(although this shortcoming is now understood to some degree [14]).  The
$qq\bar{q}\bar{q}$  interpretation of these  two states was later revived
in a different guise within the quark potential model as the ``$K\bar{K}$
molecule" interpretation [15].  Since providing a test of this particular
interpretation is one of the main results to be presented here, we first
briefly elaborate on these two models of multiquark states.

In the naive bag model the $qq\bar{q}\bar{q}$  states consist of four
quarks confined in a single spherical bag interacting via one gluon
exchange.  It is obvious that such a construction will lead to a rich
spectroscopy of states.  Although it is not clear how to treat or interpret
the problem of the stability of this spectrum under fission into two bags
[14], it is very interesting that the dynamics of this model predicts that the
lowest-lying such states will (in the SU(3) limit) form an apparently
ordinary (``cryptoexotic") nonet of scalar mesons.  It is, moreover, probable
that a better understanding of bag stability could solve both
the problem of unwanted extra predicted states and also a problem
with the $a_0$ itself: in the naive model it can ``fall apart" into
$\pi\eta$ so that it is difficult to understand its narrow width, given the
presently accepted pseudoscalar meson mixing angle (see footnote 22 in
the first of Refs. [13]).  In the absence of an understanding of
how to overcome these difficulties, we will not consider the bag picture
further in this paper\footnote{See, however, Refs. [16] for a possible
way out of the $a_0\rightarrow\pi\eta$ problem.}.

In the potential model treatment [15] it is found that the low-lying
$qq\bar{q}\bar{q}$  sector is most conveniently viewed as consisting of
weakly interacting ordinary mesons: the resulting spectrum is normally
a (distorted) two particle continuum.  Within the
ground state u,d,s meson-meson
systems, the one plausible exception to this rule is found in the
$K\bar{K}$ sector ({\it i.e.}, the
$K \bar K$ channel and those other channels strongly coupled
to it): the $L=0$ ({\it i.e.}, $J^{PC_n}=0^{++}$) spectrum seems to
have sufficient attraction to produce weakly bound states in both $I=0$
and $I=1$.  There are a number of phenomenological advantages to the
identification of these two states with the $f_0$(975) and $a_0$(980).
Among them are:

1)  It is immediately obvious why the $f_0$(975) and $a_0$(980) are
found just below $K\bar{K}$ threshold: they bear much the same
relationship to it that the deuteron bears to $np$ threshold.

2) The problem of the $f_0$ and $a_0$ widths is solved.  If these states
were $^3P_0$ quarkonia with flavours corresponding to $\omega$ and
$\rho$ (as suggested by their degeneracy), then
$\Gamma(f_0\rightarrow\pi\pi$)/$\Gamma(a_0\rightarrow\pi\eta
$) would be about 4 in contrast to the observed value of about
$\frac{1}{2}$.  At least as serious is the problem in the quarkonium
picture with the absolute widths of these states: models [17-19] predict,
for example,
\bea
\Gamma(f_0\rightarrow\pi\pi) &\simeq& (3-6) \Gamma
(b_1\rightarrow (\omega\pi)_S)\\
&\simeq & 500 -1000 MeV
\eea
versus the observed partial width of 25 MeV.  We have already noted
the problem in the bag model $qq\bar{q}\bar{q}$ interpretation with
$a_0\rightarrow\pi\eta$.  In the $K\bar{K}$ molecule picture the
narrow observed widths are a natural consequence of weak binding:
$(K\bar{K})_{I=0}\rightarrow\pi\pi$ and
$(K\bar{K})_{I=1}\rightarrow\pi\eta$  occur slowly because the
$K\bar{K}$ wavefunction is diffuse.

3) Both the $f_0$ and $a_0$ seem to bear a special relationship to
$s\bar{s}$ pairs: their $K\bar{K}$ ``couplings" are very large and they
are observed in channels which would violate the
Okubo-Zweig-Iizuka (OZI) rule [20] for an
$\omega, \rho$ -like pair of states [21].

4) The $\gamma\gamma$ couplings of the $f_0$ and $a_0$ are about
an order of magnitude smaller than expected for $^3P_0$ quarkonia [22],
but consistent with the expectations for $K\bar{K}$ molecules [23].

Although these observations argue against the viability of the $^3P_0$
quarkonium interpretation of the $f_0(975)$ (and probably also the $a_0(980)$),
they are insufficient to rule it out completely.  (Moreover, a unitarized
variant of the quark model [24], in which the scalar mesons are strongly
mixed with the meson-meson continuum, avoids several of these
problems.  In addition to this conservative alternative, the recent
analysis of Ref. [9] has raised the possibility that the $f_0$(975) is really
a combination of two effects, one of which is a candidate for a scalar
glueball.)

The main purpose of this paper is to point out a simple (and to us
unexpected) experimental test which sharply distinguishes among
these alternative explanations.  We show that the rates for
$\phi\rightarrow f_0(975)\gamma\rightarrow\pi\pi\gamma$
and $\phi\rightarrow a_0(980)\gamma\rightarrow\pi\eta\gamma$
 in the quarkonium, glueball,
and $K\bar{K}$ molecule interpretations differ significantly;
furthermore,
the ratio of branching ratios

$${{\phi\rightarrow a_0(980)\gamma}\over {\phi\rightarrow f_0(975)\gamma}}$$

\noindent also may prove to be an important datum in that it can have a
model-dependent value anywhere from zero to infinity (see Table 2)!

In the quarkonium interpretation, $\phi \rightarrow f_0(975)
\gamma$ and $\phi \rightarrow a_0(980)\gamma$ are simple electric dipole
transitions quite similar in character to several other measured
electric multipole
transitions, including not only the light quark transitions $a_2(1320)
\rightarrow \pi\gamma$, $K^*(1420)\rightarrow K\gamma$,
$a_1(1275) \rightarrow \pi\gamma$, and $b_1(1235)
\rightarrow\pi\gamma$, but also such decays as
$\chi_{c0}\rightarrow\psi\gamma$ and
$\chi_{b0}\rightarrow\Upsilon\gamma$.  From the
comparison between theory and experiment given in Ref. [17], we expect
that the quark model predictions for these processes given in Table 1 are
reliable to within a factor of two.  Thus if the $f_0$
is an $s\bar{s}$ quarkonium, the branching ratio
for $\phi \rightarrow S \gamma$ would typically be
of the order of $10^{-5}$.

  If the $f_0(975)$
is a glueball (in Ref. [9] there is a glueball component of the ``$S^*$
effect", dubbed the $S_1$(991), which couples to $\pi\pi$ and is
responsible for the resonant behaviour seen in $\pi\pi$ phase shift
analyses; the other component, dubbed the $S_2$(998), is practically
uncoupled to $\pi\pi$) then one would naturally expect
$\phi\rightarrow f_0 (975) \gamma\rightarrow\pi\pi\gamma$ to
be even smaller than in the quarkonium interpretation since the decay
would be OZI-violating.  The remarks made above on the strong decay
widths of the quarkonium states would suggest that quarkonium -
glueball mixing, through which we presume the OZI-violation would
proceed, must be small for the $f_0$ (975) to remain narrow.
Thus we can crudely estimate the glueball - quarkonium mixing angle to
be less than $[\Gamma
(f_0\rightarrow\pi\pi$)/$\Gamma(^3P_0\rightarrow\pi\pi)]^{\frac{1}
{2}}$ so that if $f_0$ (975) is a glueball
\bea
\Gamma (\phi\rightarrow f_0 \;\mbox{(glueball)}\; \gamma ) &\leq
& \frac{\Gamma
(f_0\rightarrow\pi\pi)}{\Gamma(^3P_0\rightarrow\pi\pi)}
\Gamma (\phi\rightarrow f_0\;\mbox{(quarkonium)}\; \gamma)\\
&\leq& \frac{1}{20} \Gamma (\phi\rightarrow f_0
\;\mbox{(quarkonium)}\; \gamma )
\eea
Thus if $f_0$ (975) is a glueball, this branching ratio should be more
than an order of magnitude smaller than it would be to a $\phi$-like
quarkonium.

If the $f_0$ is a quarkonium consisting only of nonstrange flavours, with
$a_0$ its isovector quarkonium partner, these states will be OZI decoupled
in the $\phi$ radiative decay.
The OZI-violating production rate via a $K\bar{K}$ loop, {\it viz.}
$\phi\rightarrow\gamma
K\bar{K}\rightarrow\gamma a_0$, may be calculated. This calculation
reveals some interesting
points of
principle which shed light on the role of finite hadron size in such loop
calculations; this calculation will be discussed in the next section.

  Interesting questions arise in the case of $qq\bar{q}\bar{q}$ or $K\bar{K}$
bound states (``molecules"). The quark contents of these two systems are
identical but their dynamical structures differ radically.
The situation here has its analog in the case of the
deuteron which
contains six quarks but is not a
``true" six-quark bound state. The essential feature
is whether the multiquark system is confined within
a hadronic system with a radius of order
 $(\Lambda_{QCD})^{-1}$
or is two identifiable colour singlets spread over a region significantly
greater than this (with radius of order $(\mu E)^{1\over 2}$
associated with the interhadron binding
energy E for a system of reduced mass $\mu$). In the
former case the branching ratio may be as large as
$10^{-4}$ (see Ref. [5] and section 4); the branching ratio for a diffuse
$K\bar K$
molecular system can be much smaller, as discussed below.

     The ratio of branching ratios is also interesting. The
ratio of $\Gamma(\phi\rightarrow\gamma a_0)/\Gamma(\phi \rightarrow
\gamma f_0)$ is approximately zero if they
are quarkonia (the $f_0$ being $s\bar{s}$ and the $a_0$ being OZI decoupled),
it is approximately unity if they are
$K\bar{K}$ systems, while
for $q^2\bar{q}^2$ the ratio is sensitively dependent on the
internal
structure of the states.
This sensitivity in $qq \bar q \bar q$ arises
because $\phi \rightarrow S \gamma$ is an $E1$ transition whose matrix
element, being proportional
to $ \Sigma e_i \vec r_i$, probes the electric charges of the
constituents weighted by their vector distance from the overall centre of mass
of the system. Thus, although the absolute transition rate
for $S=qq\bar q \bar q$ depends on unknown dynamics, the ratio
of $a_0$ to $f_0$ production will be sensitive to the internal
spatial structure of the scalar mesons through the relative
phases in $I=0$ and 1 wavefunctions and the relative spatial
distributions of quarks and antiquarks.

    For example, suppose that the state's constituents
are distributed about
the centre
of mass
with the structure $(q\bar{s})(\bar{q}s)$, where $q$ denotes $u$ or $d$,
and $(ab)$ represents a spherically symmetric cluster. Then
\bea
\left\{ \begin{array}{ll}
f_0 \\ a_0 \end{array} \right\}
&=&{1 \over {\sqrt{2}}}[(u\bar s)
(\bar us) \pm (d \bar s)(\bar ds)]
\eea
\noindent and the $E1$ matrix element will be
$$
M \sim [(e_u+e_{\bar s}) \pm (e_d+e_{\bar s})]=e_{K^+} \pm e_{K^0}
$$
and hence the ratio
$\Gamma(\phi\rightarrow\gamma f_0)/\Gamma(\phi \rightarrow
\gamma a_0)$ will be unity.
The quarks are distributed {\it as if} in a $K \bar K$ molecular
system (which is a specific example of this configuration)
 and only the
absolute
branching ratio will distinguish $q^2\bar{q}^2$ from $K\bar{K}$.

    If the distribution is
$(q\bar{q})(s\bar{s})$
then the matrix element
$$
M \sim [(e_q+e_{\bar q})-(e_s+e_{\bar s})]=0~~~.
$$
Here the quark distributions mimic $\pi ^0 \eta$ (in the $a_0$)
or $\eta \eta$ (in the $f_0$). In this case the absolute branching ratios
will be suppressed. Most interesting is the case where $S=D \bar D$,
where $D$ denotes a diquark, {\it i.e.} where
\bea
\left\{\begin{array}{ll}
f_0 \\ a_0 \end{array} \right\}&=&{1 \over {\sqrt{2}}}[(us)
(\bar u \bar s) \pm (ds)(\bar d \bar s)]
\eea
\noindent in which case
$$
M \sim [(e_u+e_s) \pm (e_d+e_s)]
$$
so that
$$
{{\Gamma(\phi \rightarrow \gamma a_0)}
\over {\Gamma(\phi \rightarrow \gamma f_0)}}=({{1+2} \over {1-2}})^2
=9~~~.
$$
The absolute rate
in this case depends on an unknown overlap
between $K \bar K$ and the diquark structure;
nonetheless the dominance
of $a_0$
over $f_0$ would be rather distinctive.
For convenience these possibilities are summarised in Table 2.

\section{The $K\bar K$ Loop Contribution to $\phi\rightarrow S\gamma$}
{}~~

     The $\phi$ and the S
(where $S=a_0$ or $f_0$) each couple strongly to $K\bar{K}$,
with the couplings
$g_\phi$ and $g$ for $\phi K^+K^-$ and $SK^+K^-$ being related to the
widths by
\be
\Gamma (\phi\rightarrow K^+K^-)=\frac{g^2_\phi }{48\pi m^2_\phi}
(m^2_\phi -
4m^2_{K^+})^{3/2}
\ee
and
\be
\Gamma (S\rightarrow K^+K^-)=\frac{g^2 }{16\pi m^2_S} (m^2_S -
4m^2_{K^+})^{1/2}
\ee

\noindent for
kinematical conditions where the decay is allowed.
Hence, independent of the dynamical nature of the S, there is an
amplitude
$M(\phi\rightarrow S\gamma)$ for the decay
$\phi\rightarrow S\gamma$ to proceed through
the charged
$K$ loop (fig. 1), $\phi\rightarrow K^+K^-\rightarrow
S(\ell)+\gamma$ where the $K^\pm$ are real or virtual and $S$ is the
scalar meson
with four momentum $\ell$.  The amplitude describing the decay can be written
\be
M(\phi (p,\epsilon_{\phi})\rightarrow S(\ell)+\gamma (q,\epsilon_{\gamma}))=
\frac{eg_\phi g}{2\pi^2im^2_K} I(a,b)[(p\cdot q)
(\epsilon_{\gamma}\cdot \epsilon_{\phi})-(p\cdot \epsilon_{\gamma})
(q\cdot \epsilon_{\phi})]
\ee
where $\epsilon_{\gamma}$ and $\epsilon_{\phi}$ ($q$ and $p$)
denote $\gamma$ and $\phi$ polarisations
(momenta).

The quantities $a,b$ are defined as $a=\frac{m^2_\phi }{m^2_K},
b=\frac{\ell^2}{m^2_K}$
so that $a-b = \frac{2p.q}{m^2_K}$ is proportional to the photon
energy, and
$I(a,b)$ which arises from the loop integral is
\be
I(a,b) = \frac{1}{2(a-b)} -\frac{2}{(a-b)^2}
\{f(\frac{1}{b})-f(\frac{1}{a})\} +\frac{a}{(a-b)^2}
\{g(\frac{1}{b})-g(\frac{1}{a})\}
\ee
where
\bea
f(x) &=& \left\{ \begin{array}{ll}
-(\arcsin (\frac{1}{2\sqrt{x}}))^2 & x>\frac{1}{4}\\
\frac{1}{4} [\ln (\frac{\eta_+}{\eta_-}) -i\pi]^2 & x<
\frac{1}{4}\end{array}\right.\nonumber
\\
g(x) &=& \left\{ \begin{array}{ll}
(4x-1)^{1/2} \arcsin (\frac{1}{2\sqrt{x}}) & x>\frac{1}{4}\\
\frac{1}{2}(1-4x)^{1/2} [\ln (\frac{\eta_+}{\eta_-}) -i\pi] & x<
\frac{1}{4}\end{array}\right.\nonumber\\
\eta_{\pm} &=& \frac{1}{2x} (1\pm (1-4x)^{1/2})
\eea

\noindent Note that $\ell^2$ may in general be virtual,
though we
shall here concentrate on the real resonance production where
$\ell^2=m_S^2$ with $m_S\simeq $
975 or 980
MeV.

Even though Refs. [1-4] use essentially the same values for the
couplings and
other parameters, they obtain different results. Our results confirm
those of Ref. [1]
apart from a minor numerical error. Ref. [5] claims that the value of the
loop
 calculation depends on the dynamical nature of the S. Since the coupling
$S\rightarrow K\bar{K}$
is input from data it is somewhat surprising that the result can discriminate
amongst models of the S. We confirm the numerical result of Ref. [5]
and discuss its physical significance below.

    The resonant contributions to
the $\phi\rightarrow K^0\bar{K}^0\gamma$ branching
fraction give
a differential decay width
\be
\frac{d\Gamma}{dk^2} = \frac{|I(a,b)|^2g^2_\varphi
g^2}{4m^4_{K}\pi^4} \chi
\ee

\noindent where $\chi$ is given by
\be
\chi = \frac{\alpha}{128\pi^2 m^3_\varphi} \frac{\frac{1}{3} (m^2_\varphi -
\ell^2)^3 (1-
\frac{4m^2_K}{\ell^2})^{1/2}}{(\ell^2-m^2_{S})^2 + m^2_{S}\Gamma^2_{S}}
\ee

\noindent Here $\ell^2$
is the invariant mass squared of the final $K^0\bar{K}^0$ system, and
hence the
resonance.

The limitations and problems in the existing literature
concerning
attempts to calculate the above are discussed in Ref. [11].
Here we shall briefly review the loop calculation in order to assess the
existing literature and to highlight the novel features of the case where
the S is a $K\bar{K}$ bound state with a finite size.

\subsection*{Calculation of the integral $I(a,b)$}
 ~~

     Upon making the $\phi$ and $K$ interactions gauge invariant, one finds for
charged
kaons
\be
H_{int} = (eA_\mu +g_\phi \phi_\mu) j^\mu - 2eg_\phi A^\mu \phi_\mu
K^{\dagger}K
\ee
where $A^\mu, \phi_\mu$ and $K$ are the photon, phi and charged kaon
fields,
$j^\mu=iK^{\dagger} (\vec{\partial}^\mu -
\\\stackrel{\leftarrow}{\partial}^\mu)
K$.  If the coupling of the kaons to the scalar meson is assumed
to be simply the {\it point-like} one $SK^+K^-$, then
gauge invariance
generates no extra diagram and the resulting diagrams are in figs. (1).
Immediately one
notes a problem: the contact diagram fig. (1a) diverges.  The trick has been to
calculate the
finite sum of figs. (1b) and (1c)
and then, by appealing to gauge invariance, to extract the correct finite
part of fig. (1a). This
is done either by

a) (Refs. [1-3]) Fig. (1a) contributes to $A^\nu \phi^\mu g_{\mu\nu}$ whereas
figs. (1b) and (1c)
contribute both to this and to $p_\nu q_\mu A^\nu \phi^\mu$.  Therefore
one
need calculate only the latter diagrams,
since the finite coefficient of the $p_\nu q_\mu$
term determines the result
by gauge invariance.

b)  (Ref. [5])  These authors
compute the imaginary part of the amplitude (which arises only from
figs. (1b) and (1c))
and write a {\it subtracted} dispersion relation, with the
subtraction constrained
by gauge
invariance. This is also sufficient to determine the amplitude.

   In section 4 we shall consider the case where the scalar meson is an
extended
object, in
particular a $K\bar{K}$ bound state.  The $SK\bar{K}$ vertex in this case
involves
a
momentum-dependent form factor $f(k)$, where $k$ is the kaon, or loop,
momentum
which will be scaled in $f(k)$ by $k_0$, the mean momentum in the bound state
wavefunction or, in effect, the inverse size of the system.  In the limit where
$R\rightarrow 0$ (or $k_0\rightarrow\infty$) we recover the formal results of
approaches
$(a,b) $ above, as we must, but our approach offers new insight
into the
physical
processes at work.  In particular, in this more physical case
there is a further diagram (fig. (2d)) proportional to
$f^\prime (k)$ since minimal substitution yields
\be
f(\vert \vec k-e\vec A \vert)-f(\vert \vec k \vert)=
-e\vec A \cdot \hat{k}\frac{\partial f}{\partial k}
\ee
As we shall see, this exactly cancels the contribution from the seagull diagram
fig. (2a) in the
limit where $q_\gamma\rightarrow 0$, and
gives an expression for the finite amplitude which is explicitly in the form of
a
difference
$M(q)-M(q=0)$. This makes contact with the subtracted dispersion relation
approach of Ref. [5].

   First let us briefly summarise the
calculation of the Feynman amplitude in the standard point-like
field theory approach,
as it has caused some problems in Refs. [2,3].  If we denote $M_{\mu\nu}
=[p_\nu q_\mu-
(p.q)g_{\mu\nu}] H(m_\phi,m_S,q)$
(see eq. (3.3)), then the tensor for fig. (3) may be written
(compare with eqs. 8 and 6 of Refs. [2] and [3], respectively)
\be
M_{\mu\nu} = egg_\phi \int \frac{d^4k}{(2\pi)^4}
\frac{(2k-p)_\mu (2k-q)_\nu}{(k^2-m_K^2)[(k-q)^2-m_K^2][(k-p)^2-m_K^2]}
\ee
We will read off the coefficient of $p_\nu q_\mu$ after combining the
denominators by the
standard Feynman trick so that
\be
M_{\mu\nu} = \frac{egg_\phi}{(2\pi)^4} 8\int^1_0 dz\int^{1-z}_0 dy
\int^\infty_{-
\infty} \frac{d^4k k_\mu k_\nu}{[(k-qy-pz)^2-c+i\epsilon]^3}
\ee
where $c\equiv m_K^2-z (1-z) m^2_\phi -zy (m^2_S-m^2_\phi)$. The
$p_\nu q_\mu$
term appears when we make the  shift
$k\rightarrow k +qy +pz$ to
obtain
\be
H=\frac{egg_\phi}{4\pi^2i} \int^1_0 dz\int^{1-z}_0 dy \;yz [m_K^2-z(1-
z)m^2_\phi - zy
(m^2_S-m^2_\phi)]^{-1}.
\ee
Note that $m^2_S <m^2_\phi$ and so one has to take care when performing the
$y$ integration.  One obtains (recall $a=m^2_\phi/m_K^2, b=m^2_S/m_K^2$)
\bea
H=\frac{egg_\phi}{4\pi^2im_K^2} \frac{1}{(a-b)} \{\int^1_0 \frac{dz}{z}
[z(1-z)-
\frac{(1-z(1-
z)a)}{(a-b)} \ln (\frac{1-z(1-z)b}{1-z(1-z)a} )]\nonumber\\
\frac{-i\pi}{(a-b)}\int^{1/\eta_-}_{1/\eta_+} (1-z(1-z)a){{dz}\over z}\}
\eea
where
$ \eta_\pm \equiv \frac{a}{2} (1\pm \rho)$ with $\rho\equiv \sqrt{1-4/a}$.
(In performing the integrals,
one must take care to note that $a>4$ whereas $b<4$ (which
causes
$\rho^2_a>0,\rho^2_b<0$)).  Our
calculation has so far only taken into account
the diagram where the $K^+$ emits the
$\gamma$;
the contribution for the $K^-$ is identical, so the total amplitude is
double that of
eq. (3.13) and therefore
in quantitative agreement with eqs. (3) and (4) of Ref. [1].
Straightforward
algebra confirms that this agrees with eqs. (9-11) of Ref. [5].

    Numerical evaluation, using $m(f_0) = 975$ MeV and
$g^2/4\pi=0.6$ GeV$^2$ leads
to
\be
\Gamma (\phi\rightarrow f_0\gamma)= 6\times  10^{-4} MeV
\ee
somewhat at variance with
the value of $8.5\times 10^{-4}$ MeV quoted in Ref. [1]
\footnote{ However, J. Pestieau, private
communication,
confirms our value.}.  Ref. [5] does not directly quote a
rate for $\phi\rightarrow
f_0\gamma$. Instead, it quotes values for $\phi\rightarrow
\gamma f_0\rightarrow\gamma\pi\pi$ (for example) and claim that this
depends upon
the $q\bar{q}$ or
$q^2\bar{q}^2$ structure of the $f_0$.  However, the differences in rate (which
vary by an
order of magnitude between $q\bar{q}$ and $q^2\bar{q}^2$ models)
{\it arise because
different
magnitudes for the $fK\bar K$
couplings have been used in the two cases}.  In the
$q^2\bar{q}^2$ model a value for $g^2 (fK \bar K)$
was used identical to ours and, if
one
assumes a unit
branching ratio for $f_0\rightarrow\pi\pi$, the rate is consistent with
our eq. (3.14)
(Ref. [5] has integrated over the resonance).
In the case of the $a_0$, Ref. [5] notes that in the $q^2\bar{q}^2$
model the
relation between $g^2 (a_0K\bar K)$ and $g^2 (a_0\pi\eta)$ implies $\Gamma
(a_0\rightarrow\pi\eta)\simeq $275 MeV.  In the $q\bar{q}$ model, in contrast,
Ref. [5] uses as
{\it input}
the experimental value of $\Gamma (a_0\rightarrow\pi\eta)\simeq$ 55 MeV
which
implies a reduced value for $g^2 (a_0\pi\eta)$ and, therefore, for $g^2
(a_0\bar{K}K)$: the
predicted
rate for $\phi\rightarrow\gamma a_0\rightarrow\gamma\pi\eta$ is
correspondingly
reduced.

Thus we believe that the apparent structure-dependence of the reaction
$\phi \rightarrow S \gamma$ claimed in Ref. [5] is
suspect. The calculation has
assumed a point-like scalar field which couples to point-like kaons
with a strength
that can
be extracted from experiment. The computation of a rate for $\phi\rightarrow
K\bar K\rightarrow\gamma S$ will depend upon this strength and cannot of itself
discriminate among models for the internal structure of the $S$.

We shall now consider the production of an extended scalar meson [11]
which is treated as
a $K\bar K$ system (based on the picture developed in Refs. [15]).

 \section{ $K\bar K$ loop production of an extended scalar meson}
 ~~

    Suppose that $K^+$ and $K^-$ with three momenta $\pm \vec{k}$ produce an
extended
scalar meson in its rest frame.  The interaction Hamiltonian $H=g\phi
(|\vec{k}|)SK^+K^-$ is in
general a function of momentum.  Now make the replacement $\vec{k}
\rightarrow
\vec{k}
 -e \vec{A}$, and expand $\phi (\vert \vec k-e\vec A \vert)$
to leading order in $e$; one then
finds a new electromagnetic contribution
\be
H_{K^+K^-f_0\gamma} =-eg\phi^\prime (k)\hat{k} \cdot \vec{A}~~.
\ee
The finite range of the interaction, which is controlled by
$\phi(k)$, implies that the currents flow over a finite distance during
the $K \bar K \rightarrow S$ transition: this current is the
``interaction current". The above current given by minimal
substitution is not unique, in the sense that the transverse part
$\vec \epsilon_{\gamma} \cdot \vec j$ cannot be determined by the requirement
of gauge invariance alone. However, it should describe the
process under consideration accurately since the radiated photon is
soft: the details of the interaction current are not important
in the soft photon regime [25].
The effect of this form factor is readily seen in time ordered perturbation
theory.
(In this section we will work in the non-relativistic
approximation. This suffices both to make our point of
principle and to provide numerically accurate estimates
for nonrelativistic $K \bar K$ bound states such as the
$f_0$ and $a_0$ in the Ref. [15] picture. In general
there are further time orderings
whose sum gives the relativistic theory; see below.)

    There are
four contributions:  ($H_{1,4}$ correspond
to figs. (2a) and (2d), while $H_{2,3}$ correspond to figs. (2b) and (2c),
where
the
$\gamma$
is emitted from the $K^+$ or $K^-$ leg). We write these (for momentum routing
see fig. (3))
\bea
H_{2,3} &=& 2egg_\phi\int d^3k \frac{\phi
(k)2\vec{\epsilon}_\gamma.\vec{k}(\vec{k}.\vec{\epsilon}_\phi
\pm\frac{1}{2}
\vec{q}.\vec{\epsilon}_\phi)}{D(E)D_1D(q\pm)}\\
H_1 &=& 2egg_\phi\int d^3k \frac{\phi
(k)\vec{\epsilon}_\gamma.\vec{\epsilon}_\phi}{D_1}\\
H_4 &=& 2egg_\phi\int d^3k \frac{\phi^\prime
(k)\vec{\epsilon}_\gamma.\hat{k}
\vec{\epsilon}_\phi .\vec{k}}{D(0)}
\eea
where
\bea
D_1 &\equiv & m_\phi -q-D(E)\nonumber \\
D(q^\pm) &\equiv& m_\phi -2E^\pm \\
D(0)&\equiv & m_\phi -2E(k)\\
D(E) &\equiv & E^{+} +E^{-}
\eea
and
where $E^\pm = E (k\pm q/2)$ with  $E(P)$ the energy of a kaon with
momentum $P$.
Note that $H_1$ is the (form-factor-modified) contact diagram and $H_4$ is the
new
contribution arising from the extended $SK\bar{K}$ vertex.

 After some manipulations their sum can be written
\be
H= 2egg_\phi\vec{\epsilon}_\gamma.\vec{\epsilon}_\phi\int d^3k
[\frac{\phi
(k)}{D_1}\{
1+\frac{\vec{k}^2 -(\vec{k}.\hat{q})^2}{D(E)} (\frac{1}{D(q^+)} +\frac{1}{D(q^-
)})\}
+\frac{\phi^\prime (k)|\vec{k}|}{3D(0)}]~~~.
\ee
If $\lim_{k^2\rightarrow\infty} (k^2\phi (k))\rightarrow 0$
\footnote{Actually, when $k\rightarrow\infty$ the relativistic
expressions of the next subsection
are needed. These show that $\phi(k)$ need only vanish
logarithmically to obtain convergence.}
we may integrate
the final
term in eq. (4.8) by parts
and obtain for it
\be
H_4 = 2 egg_\phi \vec{\epsilon}_\gamma . \vec{\epsilon}_\phi \int d^3k
\frac{\phi
(k)}{D(0)} \{-
1- \frac{\vec{k}^2-(\vec{k}.\hat{q})^2}{E(k)D(0)}\}
\ee
This is identical to the $\vec{q}\rightarrow 0$ limit of $H_1+H_2+H_3$, and
hence we
see explicitly that the $g_{\mu\nu}$ term ({\it i.e.,} the term proportional to
$\vec{\epsilon}_\gamma .\vec{\epsilon}_\phi$ as calculated above) is
effectively subtracted
at
$\vec q=0$ due to the partial integration of the $\phi^\prime (k)$
contribution,
$H_4$.

    If one has a model for $\phi(k)$ one can perform the integrals
numerically.
For the $K\bar K$ molecule, the wavefunction
\be
\psi (r) = \frac{1}{\sqrt{4\pi}} \frac{u(r)}{r}
\ee
is a solution of the Schrodinger equation
\be
\{ -\frac{1}{m_K} \frac{d^2}{dr^2} + v(r) \} u(r) = Eu(r).
\ee
One may approximate (see Ref. [23])
\be
v(r) =- 440 (MeV) \exp (-\frac{1}{2} (\frac{r}{r_0})^2)
\ee
with $r_0 = 0.57$ fm.  Equation (4.11) may be solved
numerically, giving $E=- 10$ MeV
and a $\psi(r)$ which for analytic purposes
may, as we shall see, be well-approximated
by

\be
\psi (r) = (\frac{\mu^3}{\pi})^{1/2} \exp (-\mu r) ;\quad \mu\equiv
\frac{\sqrt{3}}{2R_{K \bar K}}
\ee
where $R_{K \bar K}
\simeq 1.2 fm$ (thus $\psi (0) = 3\times 10^{-2} GeV ^{3/2}$; see
also Ref.
[23]). The momentum space wave function that is used in our computation
(see fig. (4)) is
thus taken to have
\be
{{\phi (k)} \over {\phi (0)}} = {{\mu^4}\over {(k^2+\mu^2)^2}}~~~.
\ee
The rate for $\Gamma (\phi\rightarrow S\gamma)$ is shown as a function of
$R_{K\bar{K}}$ in fig. (5).  The nonrelativistic approximation
eqs. (4.2-4.9) is
valid for
$R_{K \bar K}
\gapproxeq 0.3fm$ which is applicable to the $K\bar K$ molecule: for
$R_{K \bar K} \rightarrow 0$ the
fully relativistic formalism is required and has been included in the curve
displayed in fig.
5.  As $R_{K \bar K} \rightarrow 0$ and $\phi (k)\rightarrow 1$ we recover the
numerical result of the point-like field theory, whereas for
the specific $K\bar{K}$
molecule
wavefunction above one predicts a branching ratio of some $4\times 10^{-5}$
(width
$\simeq 10^{-4} MeV$).  This is only
$\frac{1}{5}$ of
the point-like field theory result but is larger than that expected for
the
production rate of an
$s\bar{s}$ scalar meson (see Tables 1 and 2).

\subsection*{Connection with Relativistic Field Theory}
{}~~

The nonrelativistic formalism is sufficient for
describing the radiation from a $K \bar K$ molecule. However,
it does not have the proper limit as $R_{K\bar K}\to 0$;
in this limit relativistic $K \bar K$ pairs are important in the loop
integral.
In this section we show how the relativistic formalism can be obtained
from time-ordered perturbation theory
and make contact with the relativistic field theory formalism of section 3.
The matrix elements for the time-orderings of fig. (6) are
$$
M_1^\mu ~=~ +i e g g_\phi \int {{d^3 k} \over {(2\pi)^3}}
            ~ \phi (|\vec k|) ~
            2 \varepsilon_\phi^\mu ~
         [~ - { 1 \over {2E_+2E_-(E_S+E_++E_-)}}
{}~~~~~~~~~~~~
$$
\be
{}~~~~~~~~~~~~~~~~~~~~~~~~~~~~~~~~~~~~
     + { 1 \over {2E_+2E_-(m_\phi-q-E_+-E_-)}}~ ]
\ee
where the first (second) term corresponds to fig. (6a) (fig. (6b)) and
$E_\pm$ is defined by $E_\pm = E(k\pm q/2)$.
Using $E_S=m_\phi-q$,
\be
-{1 \over {m_\phi -q +E_++E_-}}
=+{{2E_+} \over {(m_\phi-q+E_-)^2-E_+^2}}
- {1 \over {m_\phi -q+E_--E_+}}
\ee
and
\be
{1 \over {m_\phi -q -E_+-E_-}}
=+{{2E_-} \over {(m_\phi-q-E_+)^2-E_-^2}}
+ {1 \over {m_\phi -q-E_++E_-}},
\ee
we obtain
$$
M_1^\mu ~=~ +i e g g_\phi \int {{d^3 k} \over {(2\pi)^3}}
             ~ \phi (|\vec k|) ~
            2 \varepsilon_\phi^\mu ~
               [  ~ { 1 \over { 2E_- [(m_\phi - q +E_-)^2 - E_+^2] } }
{}~~~~~~~~~~~
$$
\be
{}~~~~~~~~~~~~~~~~~~~~~~~~~~~~~~~~~~~~~
                 + { 1 \over { 2E_+ [(m_\phi - q -E_+)^2 - E_-^2] } } ~].
\ee
Analogously, $M_2^\mu$, $M_3^\mu$, and $M_4^\mu$ are
$$
 M_2^\mu ~=~ M_3^\mu ~=~
               + ie g g_\phi \int {{d^3 k} \over {(2\pi)^3}}~ \phi (|\vec k|)~
                ~ 2 \varepsilon_\phi^\mu ~
                [\vec k^2 -(\vec k \cdot \hat q )^2]
{}~~~~~~~~~~~~~~~~~~~~~~~~~~~~~~~~~~~
$$
$$
{}~~~~~~~
               \times ~[  ~ { 1 \over { 2E_+ [(q +E_+)^2 - E_-^2]
                                    [(m_\phi+E_+)^2 - E_+^2] } }
$$
$$
{}~~~~~~~~~~~~~
                 ~+~ { 1 \over { 2E_- [(q -E_-)^2 - E_+^2]
                                    [(m_\phi-q+E_-)^2 - E_+^2] } }
$$
\be
{}~~~~~~~~~~~~~~~
                 ~+~ { 1 \over { 2E_+ [(m_\phi -E_+)^2 - E_+^2]
                                    [(m_\phi-q-E_+)^2 - E_-^2] } }
                                                                     ~ ]
\ee
$$
 M_4 ^\mu~=~ + ie g g_\phi \int {{d^3 k} \over {(2\pi)^3}} ~\phi '(|\vec k|)~
                 {{|\vec k|} \over 3 }~
                ~ 2 \varepsilon _\phi^\mu~
               [ ~  { 1 \over { 2E_0 [(m_\phi +E_0)^2 - E_0^2] } }
{}~~~~~~~~~~~~~~~~
$$
\be
{}~~~~~~~~~~~~~~~~~~~~~~~~~~~~~~~~~
                 + { 1 \over { 2E_0 [(m_\phi -E_0)^2 - E_0^2] } } ~]
\ee
where $E_0$ is defined by $E_0=E(k)$.

In this way, we obtain ``relativistic'' expressions
for the radiative $\phi$ meson decays.
Matrix elements for the process a$-$d in fig. (2)
may thus be written
\be
M_1^\mu ~=~ - e g g_\phi \int {{d^4 k} \over {(2\pi)^4}}
             ~ \phi (|\vec k|) ~
              {{2 \varepsilon _\phi^\mu} \over {D(k-q/2)D(k+q/2-p)}}
{}~~~~~~~~~~~~~~~~~~
\ee
\be
M_2^\mu ~=~ + e g g_\phi \int {{d^4 k} \over {(2\pi)^4}}
              ~\phi (|\vec k|) ~
              {{\varepsilon _\phi \cdot (2k+q-p) ~(2k)^\mu}
                \over {D(k+q/2)D(k-q/2)D(k+q/2-p)}}
\ee
\be
M_3^\mu ~=~ + e g g_\phi \int {{d^4 k} \over {(2\pi)^4}}
            ~  \phi (|\vec k|) ~
              {{\varepsilon _\phi \cdot (2k-q+p)~ (2k)^\mu}
                \over {D(k+q/2)D(k-q/2)D(k-q/2+p)}}
\ee
\be
M_4^\mu ~=~ + e g g_\phi \int {{d^4 k} \over {(2\pi)^4}}
             ~ \phi ' (|\vec k|) ~
              {{\varepsilon _\phi \cdot (2k-p) ~\hat k ^\mu}
                \over {D(k)D(k-p)}}
{}~~~~~~~~~~~~~~~~~~~~~~~~~~~~~~~~~~
\ee
where $D(k)$ is defined by
\be
D(k)~=~ k^2 - m_{_{K}} ^2
\ee
and $\hat k=(0,\vec k/|\vec k|)$.
In the particular case where $\phi (|\vec k|)=1$ and
$\phi ' (|\vec k|)=0$, these reproduce the familiar
field theory expressions of Refs. [1-5] and section 3.
It is interesting to note the role that
$\phi ' (|\vec k|)$ plays in regularising the
infinite integral.

Define the matrix elements $\tilde M_j$ ($j=1-4$)
by $\tilde M_j = \varepsilon_\gamma \cdot M_j
 /[ie\varepsilon_\gamma \cdot \varepsilon_\phi]$
and the decay width is then calculated by
\be
\Gamma (\phi \to S \gamma) ~=~
      {{\alpha q} \over {3 m_\phi^2}}
   ~   | \tilde M |^2
{}~~~~,~~~~
\tilde M  = \tilde M_1 +
              \tilde M_2 +
              \tilde M_3 +
              \tilde M_4
\ee
which reproduces the expressions in Refs. [1-5] and
provides a check on our formalism.
Eqs. (4.21-4.24), when evaluated numerically, give the decay widths shown in
Fig. 5. In the limit $R_{K \bar K} \rightarrow 0$ our numerical results agree
with eq. (3.14) which was obtained by using the point-like field theory.

\section{A Comment on the OZI Rule}
 ~~~

     The calculations presented in this paper may have a bearing on one of
the least understood characteristics of the low energy strong interactions:
the Okubo-Zweig-Iizuka (OZI) rule [20]. If the
$a_0$ were a $\frac{1}{\sqrt{2}} (u\bar{u}-d\bar{d})$ state,
its production in
$\phi\rightarrow
a_0\gamma$ would vanish in ``lowest order" in the quark model,
with the $K \bar K$ loop contribution presumed to provide
a small correction since such processes are OZI-violating ({\it e.g.,}
$\omega-\phi$ mixing could also occur {\it via} such loops).
We have seen that in the point-like approximation
$\phi\rightarrow
a_0\gamma$ would proceed with a
branching ratio of order $10^{-4}$ via this loop process, as would $f_0
=\frac{1}{\sqrt{2}}
(u\bar{u}+d\bar{d})$. If $f_0 =s\bar{s}$, a similar rate
would be obtained from the $K \bar K$ loop, but now there would
be a direct term which is supposed to be dominant.
It is, however, easy to discover that this direct process would only
produce a branching ratio
of the order of $10^{-5}$ (see Table 1).

     Our calculation provides some insight into this conundrum. If
the $K \bar K$ system is diffuse,
$R_{K \bar K} \gapproxeq 2fm$, then the loop calculation gives a branching
ratio
$<10^{-5}$ (see fig. (5))
and the empirical OZI rule is good.  Physically, the rate
is suppressed due to the
poor
spatial overlap between the $K\bar{K}$ system and the $\phi$.
The point-like field theory does not allow for this:
superficially the loops have a
large
magnitude.
The essential observation is that the point-like calculation does not
take into account the confinement scale, even though it is clear
from our results that the dynamics can depend on it rather critically.

   Now consider a $\phi$ and
assume that $S$ is an $ (s\bar{s})$ scalar meson,
confined in $\Lambda^{-1}_{QCD}
\simeq 1~fm$ and connected by an intermediate state
with quark composition $q\bar q s \bar s$. If this multiquark
system is confined in a length scale
$\lapproxeq  \Lambda^{-1}_{QCD}  \simeq
1fm$ ({\it i.e.}, it
is a ``genuine" $q^2\bar{q}^2$ state
and separate identifiable kaons are not
present), then the point-like field theory calculations, which contain no
intrinsic length,
are superficially at least roughly applicable. The
$\phi\rightarrow\gamma S$ branching ratio
{\it via} the $K \bar K$ part of this compact system
is then elevated above the $10^{-
5}$ barrier.
However, if a pure $K\bar{K}$ intermediate state forms, then it must
occupy $>2\Lambda_{QCD}^{-1}$. The amplitude for the $\phi$ or a $S(s\bar{s})$
to fluctuate to this scale of size
would be small and it is this supression
that is at the root of the OZI rule in this process.

      We see from this reasoning that the contribution of
diagrams which correspond at the quark level
to $q \bar q s \bar s$ loops really contain two distinct contributions
at the hadronic level. These are first of all the diffuse
contributions which can arise from hadronic loops corresponding
to nearby thresholds, in this case from $K \bar K$. Then there
are ``short distance" contributions where approximating the
$q\bar q s \bar s$ system as a $K \bar K$ system is potentially
very misleading: a realistic calculation of such contributions would at least
have to include a very large set of hadronic loops.
A step in this direction has recently been taken in Refs. [26].
These authors have considered the loop contributions to, {\it e.g.},
$\omega-\phi$ mixing in the $^3P_0$ quark pair creation model, and
found that there is a systematic tendency for the sum of all
hadronic loops to cancel. In fact, they show that (in their model)
the incompleteness of the cancellation  of OZI-violating
hadronic loops is precisely
due to nearby thresholds.

\section{Conclusions}
{}~~

There is still much thought needed on the correct modelling of the $K\bar K$ or
$q^2\bar{q}^2$
scalar meson and the resulting rate for
$\phi\rightarrow S\gamma$: the present paper merely makes a start by
clarifying the present literature, making the first predictions for the
production of a $K\bar K$ molecule, and pointing out the utility of the ratio
of branching ratios as a filter. However, these results in turn raise questions
that merit further study. For example, there
 are interesting interference effects possible
between the $a_0 (I=1)$ and $f_0 (I=0)$ states which have not been
considered. These two nearly degenerate states lie so near to
 the $K \bar K$ thresholds
that the mass difference between neutral and charged kaons
is not negligible: for example,
their widths straddle the $K^+ K^-$ threshold
but only barely cross the $K^0 \bar{K^0}$ threshold (at least in the case
of the relatively narrow $f_0$).

    Although there is clearly much to be done, it is already
clear that
there may be unique opportunities for
probing dynamics in $\phi\rightarrow S\gamma$ and investigating
the nature of the scalar mesons below 1 GeV.
Moreover, we can already
conclude that the branching ratio of $\phi \rightarrow S \gamma$ will be
between
$10^{-4}$ and $10^{-5}$ depending on the dynamical nature of these scalars
and so will generate nugatory\footnote{ {\bf n$\bar u$g' atory}, a. Trifling,
worthless, futile; inoperative, not valid. [f. L {\it nugatorius
(nuggari} trifle f. prec., -ORY)] ~~~[28]}
 background to studies of CP-violation
at DA$\Phi$NE or other $\phi$-factories.

\section*{Acknowledgements}

{}~~

We are indebted to the Institute for Nuclear Theory in Seattle
 for their hospitality.
F.E.C. would like to dedicate this paper to the memory
of Nick Brown, who was very interested in this physics.
He would also like to thank D. Ross
and P. Valovisky for comments, the organisers of
the
DA$\Phi$NE  workshops, and M. Pennington and G. Preparata for their
interest in the
OZI rule and some technical aspects of this work. N.I. would
like to thank the Department of Theoretical
Physics of the University of Oxford for its hospitality
during the period when this work was begun and, with F.E.C., to express
his gratitude to the referee who rejected an earlier incorrect
version of this paper [27].
S.K. acknowledges the support of the U.S. NSF under contract
NSF-PHY91-08036; N.I. was supported during the period this
work was done by NSERC Canada and by U.S. DOE contract
DOE-AC05-84ER40150.

\newpage
\begin{thefiglist}{10}
\item{ } The contact (a) and loop radiation (b,c) contributions.
\item{ } As fig. 1 but with an extended scalar meson.  Note the new diagram
(d).
\item{ } Momentum routing.
\item{ } Comparison between the exact momentum space wavefunction
$\phi (k)$ (solid) and the approximation
of eq.(4.14); $k$
is the relative momentum of the $K$ and $\bar K$.
\item{ } $\Gamma (\phi\rightarrow S\gamma)$ in MeV versus $R_{K\bar{K}}$
in $fm$.
\item{ } The two time orderings of fig. 2(a).
\end{thefiglist}

\end{document}